\documentclass[a4paper,11pt]{amsart}

\usepackage{url}
\usepackage{cite}

\usepackage{amssymb}
\usepackage{color}

\usepackage[normalem]{ulem}
\usepackage[utf8]{inputenc}
\usepackage{graphicx}

\usepackage{soul}

 \newcommand{\griemg}{{\griem_{[\gamma]}}{}}

\newcommand{\wmcO}{\,\,\widehat{\!\!\mcO}}
\newcommand{\mcU}{{\mycal U}}
\newcommand{\mcO}{{\mycal O}}

\newcommand{\qedskip}{\hfill $\Box$ \medskip}

\newcommand{\zP}{{\mathring{P}}}

\newcommand{\znabla}{{\mathring\nabla}}

\newcommand{\bean}{\begin{eqnarray}\nn}

\newcommand{\nn}{\nonumber}

 \newcommand{\mcV}{{\mycal V}}

\newcommand{\zg }{{ \mathring{g}}}
\newcommand{\zV }{{ \mathring{V}}}
\newcommand{\zglorentz}{{ \mathring{\glorentz}}}
\newcommand{\glorentz}{{ {\mathbf g}}}

\newcommand{\zDelta}{{ {\mathring \Delta}}}
\newcommand{\zRriem}{{ {\mathring \Rriem}}}
\newcommand{\Rric}{{ {\mathfrak{Ric}}}}
\newcommand{\zRric}{{ \mathring{\mathfrak{Ric}}}}

\newcommand{\Rriem}{{ {\mathfrak R}}}

\newcommand{\RGamma}{{ {\mathfrak G}}}
\newcommand{\griem}{{ {\mathfrak g}}}

\newcommand{\zgriem}{{ \mathring{\mathfrak g}}}

\DeclareFontFamily{OT1}{rsfs}{}
\DeclareFontShape{OT1}{rsfs}{m}{n}{ <-7> rsfs5 <7-10> rsfs7 <10->
rsfs10}{} \DeclareMathAlphabet{\mathscr}{OT1}{rsfs}{m}{n}

\newcommand{\bel}[1]{\begin{equation}\label{#1}}
\newcommand{\beal}[1]{\begin{eqnarray}\label{#1}}
\newcommand{\beadl}[1]{\begin{deqarr}\label{#1}}
\newcommand{\eeadl}[1]{\arrlabel{#1}\end{deqarr}}
\newcommand{\eeal}[1]{\label{#1}\end{eqnarray}}
\newcommand{\eead}[1]{\end{deqarr}}
\newcommand{\eea}{\end{eqnarray}}
\newcommand{\eeaa}{\end{eqnarray*}}

\newcommand{\be}{\begin{equation}}
\newcommand{\ee}{\end{equation}}

\DeclareFontFamily{OT1}{rsfs}{}
\DeclareFontShape{OT1}{rsfs}{m}{n}{ <-7> rsfs5 <7-10> rsfs7 <10->
rsfs10}{} \DeclareMathAlphabet{\mycal}{OT1}{rsfs}{m}{n}

\newcommand{\mcL}{{\mycal L}}

\newcounter{mnotecount}[section]

\renewcommand{\themnotecount}{\thesection.\arabic{mnotecount}}

\newcommand{\N}{{\mathbb N}}
\newcommand{\Z}{{\mathbb Z}}
\newcommand{\mnote}[1]
{\protect{\stepcounter{mnotecount}}$^{\mbox{\footnotesize $
\bullet$\themnotecount}}$ \marginpar{
\raggedright\tiny\em $\!\!\!\!\!\!\,\bullet$\themnotecount: #1} }

\newcommand{\rmnote}[1]{}

\def\mysavedown#1{\edef\mysubs{\mysubs#1}}
\def\mysaveup#1{\edef\mysups{\mysups#1}}
\def\mydown#1{{\mytensor}_{\vphantom{\mysubs}#1}}
\def\myup#1{{\mytensor}^{\vphantom{\mysups}#1}}
\def\tensor#1#2{
  #1
  \def\mytensor{\vphantom{#1}}
  \def\mysubs{\relax}
  \def\mysups{\relax}
  \let\down=\mysavedown
  \let\up=\mysaveup
  #2
  \let\down=\mydown
  \let\up=\myup
  #2
  }

\newcommand{\C}{\mathbb C}
\newcommand{\R}{\mathbb R}

\renewcommand{\to}{\rightarrow}

\renewcommand{\epsilon}{\varepsilon}
\renewcommand{\hat}{\widehat}

\def\crn#1#2{{\vcenter{\vbox{
        \hbox{\kern#2pt \vrule width.#2pt height#1pt
           }
          \hrule height.#2pt}}}}

\renewcommand{\hbar}{{\overline h}}

\newcommand{\pre}[2]{{{\vphantom{#2}}^{#1}}\kern-.2ex{#2}}

\theoremstyle{plain}
\newtheorem{theorem}{\sc Theorem}[section]

\newtheorem{Proposition}[theorem]{\sc Proposition}

\theoremstyle{definition}
\newtheorem{definition}[theorem]{Definition}

\newtheorem{Remark}[theorem]{\sc  Remark\rm}

\numberwithin{equation}{section}

\date{\today}

\renewcommand{\mnote}[1]{}

\begin{document}

\title[Periodic solutions of nonlinear wave equations] {On periodic solutions of nonlinear wave equations, including  Einstein equations with a negative cosmological constant}
\thanks{Preprint UWThPh-2017-41}

\author[P.T. Chru\'sciel]{Piotr T.~Chru\'sciel}

\address{Piotr
T.~Chru\'sciel, Faculty of Physics,
 University of Vienna, Boltzmanngasse 5, A1090 Wien, Austria}
\email{piotr.Chru\'sciel@univie.ac.at} \urladdr{http://homepage.univie.ac.at/piotr.chrusciel/}

\begin{abstract}
We construct periodic solutions of nonlinear wave equations using analytic continuation. The construction applies in particular to Einstein equations, leading to infinite-dimensional families of time-periodic solutions of the vacuum, or of the
Einstein-Maxwell-dilaton-scalar fields-Yang-Mills-Higgs-Chern-Simons-$f(R)$
equations, with a negative cosmological constant.
\end{abstract}

\maketitle

\tableofcontents

\section{Introduction}
 \label{section:intro}

Time-periodic solutions of nonlinear hyperbolic partial differential equations are notoriously difficult to construct (cf., e.g., \cite{CraigWayne,Wayne} and references therein). A systematic construction of such solutions is bound to be particularly delicate for Einstein equations, where various no-go theorems are available~\cite{galloway-splitting,BicakScholtzTod,AlexakisSchlue}.
Thus, establishing existence of large class of such solutions appears to be a particularly challenging task.

It came therefore as a surprise when, in a remarkable recent paper, Maliborski and Rostworowski provided numerical evidence for existence of families of periodic solutions of the Einstein-scalar field equations with a negative cosmological constant~\cite{MaliborskiRostworowski0} (compare~\cite{DiasHMS,Rostworowski:2017}).
Their solutions were driven by  periodic eigenmodes of the scalar field equation.
The object of this work is to prove existence  of a closely related large class of time-periodic solutions of the Einstein equations, with or without sources; see Theorem~\ref{T19XI17.1} and Remark~\ref{R19XI17.1} below. Our solutions are based on a different mechanism: they are perturbations of anti-de Sitter space-time driven by carefully-chosen free data, periodic in time, at the conformal boundary, see Definition~\ref{D13XI17.1} for the definition of the class of admissible boundary data.

The proof of existence of such solutions proceeds by analytic continuation from solutions of elliptic boundary value problems. The method involves complex-valued tensor fields in intermediate steps, seems to have been unnoticed in the literature so far, and has interest in its own. Indeed, it can be applied to other nonlinear wave equations: as a warm-up we   construct periodic solutions of  a boundary-value problem for a large class of nonlinear scalar wave equations.  The argument is inspired by a method introduced in~\cite{ChDelayKlingerBH},
where families of complex valued tensor fields were used to construct stationary black hole solutions of the Einstein equations.

\section{Toy model: non-linear scalar field on a cylinder}
 \label{s28X17.1}

The following simple model illustrates well our construction. Let
\begin{equation}\label{28X17.2}
  \glorentz= -V^2 dt^2 + g
\end{equation}
be a time-independent Lorentzian metric on $M:=\R\times \Omega$, where the closure $\overline\Omega$ of $\Omega$ is a smooth compact manifold with boundary $\partial \Omega$, $g$ is a Riemannian metric on $\Omega$ which extends smoothly across $\partial \Omega$ preserving signature, and $V$ is a strictly positive smooth function on $\bar \Omega$. (It will become clear shortly that the construction below works for a class of time-periodic, real-analytic in $t$, Lorentzian metrics, but the reader might find it conceptually simpler to consider static metrics to start with.) We wish to find time-periodic solutions of the following scalar field equation
\bel{28X17.1}
 \Box_\glorentz \phi = F(\phi)
 \quad
 \Longleftrightarrow
 \quad
 - V^{-2} \partial_t^2 \phi + \Delta_g \phi + V^{-1}g(D  V,D \phi) = F(\phi)
 \,,
\ee
for a real-valued scalar field $\phi$, where $\Delta_g$ is the Laplace operator associated with the Riemannian metric $g$. We will assume that $F$ is real-analytic, and real-valued for $\phi \in \R$.
 We assume that $\phi\equiv 0$ is a solution; equivalently,
\begin{equation}\label{5XI17.2}
  F(0)=0
  \,.
\end{equation}
Suppose that we are given a time-periodic real-valued function
$$
 \psi^0:\R\times \partial \Omega\equiv \partial M \to \C
 \,.
$$
We raise the question, is there a solution of \eqref{28X17.1} which coincides with $\psi^0$ on $\partial M$?

Note that there are many such solutions which are \emph{not} periodic, obtained by solving an initial-boundary value problem for \eqref{28X17.1} with boundary data $\psi^0$ and varying the Cauchy data at $\{t=0\}\times \Omega$. (Many of those solutions might fail to exist globally on $M$, which is irrelevant for the discussion that follows.) So the issue is, do there exist Cauchy data on $\Omega$ which would lead to periodic solutions with the right boundary values on $\partial M= \R\times \partial \Omega$? (Such solutions will of course exist globally.)

We do not know the answer to this question for general $\psi^0$. However, we will show existence of a large class of \emph{real analytic in time and  time-periodic} boundary functions $\psi^0$ for which the problem can be solved. This will be done using analytic continuation.

In order to understand how analyticity might be relevant, let us introduce the Riemannian metric
\begin{equation}\label{28X17.4}
  \griem = V^2 d\varphi^2 + g
 \,,
\end{equation}
where $\varphi$ is a periodic coordinate on $S^1$, with period equal to, say $1$. We replace \eqref{28X17.1}
by
\begin{equation}\label{28X17.6}
  \Delta_\griem \phi = F(\phi)
 \quad
    \Longleftrightarrow
 \quad
  V^{-2} \partial_\varphi^2 \phi + \Delta_g   \phi + V^{-1}g(DV, D \phi )= F(\phi)
 \,,
\end{equation}
where $\phi$ is allowed to be complex valued. Suppose that we have a family of solutions of \eqref{28X17.6} parameterised by a parameter $t$,
$$
 t\mapsto \phi_t(\varphi,x)\,,
 \quad
  x\in \Omega
 \,,
$$
such that the family of functions
$$
 \phi(t+i\varphi,x)\equiv\phi(t,\varphi,x):=\phi_t(\varphi,x)
 \,,
$$
with $(t,\varphi)$ in an open subset of $\C$, is \emph{analytic in the complex variable}
\begin{equation}\label{28X17.7}
  \tau:= t+ i \varphi
 \,.
\end{equation}
The Cauchy-Riemann equations give
\begin{equation}\label{28X17.8}
  \partial_\varphi \phi = - i \partial_t \phi
 \,,
\end{equation}
from which it follows that for \emph{every fixed value of $\varphi$} the function
\begin{equation}\label{28X17.9}
  t \mapsto \phi^\varphi(t,x):=\phi(t+i \varphi ,x)
  \,,
\end{equation}
with $x\in \Omega$, solves \eqref{28X17.2}.

In what follows, functions which are periodic in a variable $s\in \R$, or $\varphi \in \R$, will be identified with functions on $S^1$, and vice-versa.

The idea of the construction is now the following: Choose a one-parameter family of boundary data $\psi_t(\varphi,x):=\psi(t+ i \varphi,x )$, defined for $x\in \partial \Omega$ and parameterised by $t$, for the elliptic boundary value problem
\begin{equation}\label{28X17.6a}
 \Delta_\griem
\phi_t = F(\phi_t)
 \,,
  \quad
  \phi_t (\varphi,x)= \psi_t(\varphi, x)
   \ \mbox{for} \ (\varphi, x) \in S^1\times \partial\Omega
   \,,
\end{equation}
arising from a  function $\psi(t+ i \varphi , x)$ which is holomorphic in $\tau=t + i \varphi$, thus
 $\partial_{\overline \tau} \psi =0$, and  doubly-periodic in the following sense:
\begin{equation}
 \label{5XI17.1}
  \psi(t+ i \varphi , x)
  =\psi(t + 1 + i \varphi, x)
  =\psi(t + i(\varphi+1), x)
  \,,
\end{equation}
for all $(t,\varphi,x)$ in the domain of definition of $\psi$.
Suppose that the linearisation of \eqref{28X17.6},
\begin{equation}\label{5XI17.4}
  P(\phi):= 
 \Delta_\griem   - F'(\phi)
\end{equation}
at $\phi=0$ is an isomorphism on the space of functions with vanishing data on $S^1 \times  \partial \Omega$.
We can then use the implicit function theorem to solve \eqref{28X17.6} with initial values $\epsilon \psi_t (\varphi,x):= \psi(t+i \varphi, x)  $, for all $\epsilon $ small enough and $t$ near $0$. Then
\begin{equation}\label{5XI17.5}
  0=\partial_{\overline \tau} \Big(
 \Delta_\griem
 \phi_t - F(\phi_t)\Big)
   = P(\phi_t) \partial_{\overline \tau} \phi_t
   \,.
\end{equation}
Since $P$ has no Dirichlet kernel for $\phi_t$ sufficiently small and since $\partial_{\overline \tau} \psi_t =0$ on $S^1\times \partial \Omega$, we find that
\begin{equation}\label{5XI17.6}
   \partial_{\overline \tau} \phi_t =0
   \,.
\end{equation}
Hence the function $\phi(t+i\phi,x):= \phi_t(\phi,x)$ is holomorphic in $\tau$.

So far we have only defined $\phi(t+i\varphi,x)$ for small $t$. For each $x$ let us denote by the same symbol $\phi(\tau,x)$ an analytic extension, in $\tau$, of the function defined so far. 
By definition, \eqref{28X17.6a} thus holds for all open sets of pairs $(\tau,x)$ for which the extension exists.

\emph{If}
\begin{enumerate}
  \item[P1.]  for all $x$ the functions $\phi(\cdot, x)$ can be extended from the imaginary axis to a subset of $\C$ containing the interval $[0,1]$, and
  \item[P2.] \emph{if} the functions $t\mapsto \phi(t+i0,x)$ are periodic in $t$, and
  \item[P3.] \emph{if} these functions are real-valued,
\end{enumerate}
we will have obtained the desired periodic real-valued solution.

\subsection{On P1.}
In order to justify P1 we introduce the class of differentiable curves $s\mapsto \gamma(s) = t(s) + i \varphi(s)$ such that $t(s+1)=t$ $\varphi(0)=0 $, $\varphi(s+1)=\varphi(s)+1$, with $\partial_s\varphi>0$. Such curves will be called \emph{admissible}. Consider an admissible curve such that $\gamma(s)$ lies within the domain of definition of $\phi(\tau,x)$ for all $x$. Along this curve we have, using the Cauchy-Riemann equations,
\begin{equation}\label{5XI17.11}
  \partial_s \phi
   = \partial_t \phi \frac{\partial t}{\partial s}
   +\partial_\varphi \phi \frac{\partial \varphi}{\partial s}
   = -i \partial_\varphi \phi \frac{\partial \gamma}{\partial s}
    \qquad
    \Longleftrightarrow
    \qquad
      \partial_\varphi \phi
 =
  i \frac{   \partial_s \phi}
   {     \partial_s \gamma
    }
   \,.
\end{equation}
It follows thus from \eqref{28X17.6} that the function
$$
 (s,x) \mapsto \phi_{[\gamma]}(s,x):= \phi(\gamma(s),x)
$$
is a solution of the boundary value problem
\begin{equation}\label{28X17.6b}
 \big((V\partial_s \gamma)^{-2} \partial_s^2 + \Delta_g\big) \phi = F(\phi)
 \,,
  \quad
  \phi_{[\gamma]} (s,x)= \psi(\gamma(s), x)
   \ \mbox{for} \ x \in \partial\Omega
   \,,
\end{equation}
We wish to analyse when this problem is elliptic. For this consider the (possibly complex-valued) symbol $\sigma$ of the operator appearing at the right-hand side of \eqref{28X17.6b}: setting $k= (k_s,k_i)$ we have
\begin{equation}\label{5XI17.21}
  \sigma(k)=(V\partial_s \gamma)^{-2} k_s^2 + g^{ij}k_i k_j
  \,.
\end{equation}
Rewritten in terms of the real and imaginary part of $\phi= \Re \phi + i \Im \phi$, the equation $\sigma(k) \phi=0$ reads
\begin{equation}\label{5XI17.22}
  \left(\begin{array}{cc}
          \Re\big((V\partial_s \gamma)^{-2}\big) k_s^2 + g^{ij}k_i k_j & - \Im \big((V\partial_s \gamma)^{-2}\big) k_s^2 \\
           \Im \big( (V\partial_s \gamma)^{-2}) \big) k_s^2 & \Re\big((V\partial_s \gamma)^{-2}\big) k_s^2  + g^{ij}k_i k_j
        \end{array}
        \right)
         \left(
         \begin{array}{c}
           \Re \phi \\
           \Im \phi
         \end{array}
         \right)
         =0
        \,.
\end{equation}
This has solutions $k\ne 0$ if and only if $\Im (\partial_s \gamma)^{-2}=0$ and $\Re(\partial_s \gamma)^{-2}\le 0$. (This could of course have been inferred directly from \eqref{5XI17.21}, without decomposing into real and imaginary parts.) We conclude that
\emph{the problem \eqref{28X17.6b} is elliptic if and only if $\partial_s\varphi$ is nowhere zero on $\gamma$.}
(We also see again that we obtain a wave equation along any horizontal segment of $\gamma$.

Now, doubly-periodic, say meromorphic, functions on the plane are necessarily singular; see Appendix~\ref{A6XI17.1} for some remarks on such functions. When $\psi$ is has no singularities near the axes, we can use the above observation to claim that:

\begin{Proposition}
 \label{P5XI17.1}
 Suppose that there exists an open neighborhood $\mcU \subset \C$ of $
i\R$  in $\C$ such that $\psi$ is smooth on $\mcU\times\partial \Omega$, and that $\psi(\cdot, x)$ is analytic on $\mcU$ for all $x\in \partial \Omega$.
Consider an open neighborhood $\mcO\subset \overline \mcO \subset \mcU$ of $i\R$  which is covered by a family of admissible curves $\gamma_\lambda$, $\lambda \in [-1,1]$, such that the map
\begin{equation}\label{20XI17.11}
 (-1,1) \times \R \ni (\lambda,s) \mapsto (\gamma_\lambda(s),s)
\end{equation}
is a diffeomorphism.
 If the linearisation of \eqref{28X17.6b} with respect to $\phi$ at $\phi=0$, with $\gamma$ replaced by $\gamma_\lambda$, is invertible for all $\lambda\in [-1,1]$, then  there exists $\epsilon_0 >0$ such that the functions $\phi(\tau,x)$ are holomorphic in $\tau$ and defined on $\mcO \times \bar \Omega$.
\end{Proposition}

\noindent{\sc Proof:}
There exists $\epsilon_0>0$ so that for all $\lambda \in [-1,1]$ and $0 < \epsilon \le \epsilon_0$ we can solve the boundary value problem
\begin{equation}\label{28X17.6c}
\left\{
  \begin{array}{ll}
 \big((V\partial_s \gamma)^{-2} \partial_s^2 + \Delta_g\big) \phi + V^{-1} g(DV,D\phi)= F(\phi)
 \,, & \hbox{$(s,x)\in S^1\times \Omega$;} \\
  \phi_{[\gamma]} (s,x)= \epsilon\psi(\gamma(s), x)
 \,,
    & \hbox{$ x \in \partial\Omega$.}
  \end{array}
\right.
\end{equation}
Uniqueness gives $\phi_{[\gamma]} (s,x)= \phi(\gamma(s),x)$ for $x\in \Omega$, and the result easily follows by invoking uniqueness of analytic extensions.
\qedskip

From what has been said we see that the isomorphism property of the linearised map, required above, will hold if all the derivatives $\partial_s\gamma_\lambda-i$ are sufficiently small.

\subsection{On P2.}
It is conceivable that all the solutions provided by Proposition~\ref{P5XI17.1} will be $1$-periodic in $t$, but we have not been able to justify this in general. However, the desired periodicity becomes apparent if one moreover assumes that

\begin{enumerate}
\item 
 all singularities of $\psi$ are located on the lines $\Re z \in \frac 12 + \Z$, and that
    \item
     $\psi$ satisfies
%
\begin{equation}\label{5XI17.8}
  \psi(z) = \psi(-z)
 \,.
\end{equation}
\end{enumerate}

Uniqueness of solutions and \eqref{5XI17.8} imply that
\begin{equation}\label{5XI17.8a}
  \phi(z,x) = \phi(-z,x)
 \,.
\end{equation}

\begin{figure}[ht]
	\centering
  \includegraphics[width=0.5\textwidth]{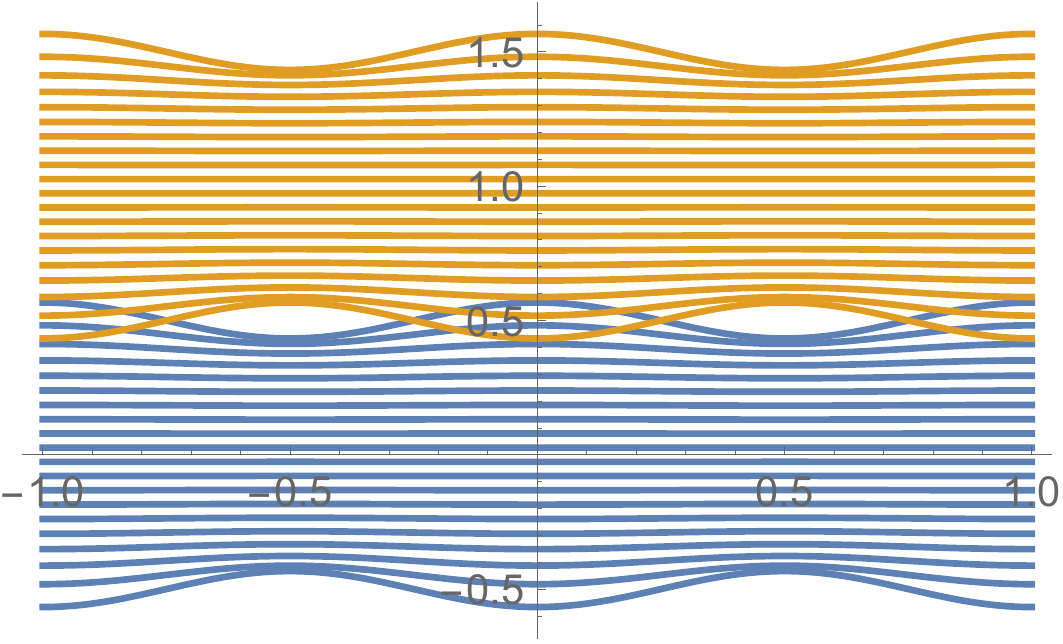}
	\caption{The image of the strips $|\Im z |\le 1/2$, and $|\Im z -1|\le 1/2$, by the biholomorphic map \eqref{11XII17.1}. The map $\chi$ of \eqref{11XII17.1+} is obtained by rotating this picture anti-clockwise by $\pi/2$. }
	\label{fig1}
\end{figure}

To see how this implies periodicity, it is convenient to choose the map in \eqref{20XI17.11}, which we will refer to as $\chi$, to be holomorphic in $\lambda +i s$. An example is given by rotating by 90 degrees the function
\begin{equation}\label{11XII17.1}
  f(z) =   z + \eta \sin (2 \pi z)
 \,,
\end{equation}
namely
\begin{equation}\label{11XII17.1+}
  \chi(\lambda + i s) = i f (s-i\lambda)
 \,,
\end{equation}
see Figure~\ref{fig1}. Here $\eta$ is a small real positive parameter. The curves $\gamma_{\lambda}$ in \eqref{20XI17.11} are then defined as $\gamma_\lambda(s) = \chi(\lambda +is)$.
Note that $f(z)=-f(-z)$. One can choose $\eta$ small enough so that $f$ is a bijection from $\R\times [-1/2,1/2]$ to its image, with all singularities of $\psi$ lying outside the image  $\cup_{\lambda,s} \{\gamma_\lambda (s)\}$ of $[-1/2,1/2]\times \R$ by the map $\chi$.

As in Proposition~\ref{P5XI17.1}, for $\lambda\in [-1/2,1/2]$ we consider the associated family
$$
 \gamma_\lambda (s)= t_\lambda(s) + i \varphi_\lambda(s)
 \,,
$$
with $\mcO$ defined as the image of the strip $(-1/2,1/2)\times \R\subset \C$. We set
\begin{equation}\label{5XI18.9}
  \hat \gamma_\lambda = 1 - t_\lambda(s) - i \varphi_\lambda (s)
  \,.
\end{equation}
Then the family of curves $\{\hat \gamma_\lambda\}_{\lambda \in (-1/2,1/2)}$ covers a neighborhood $\wmcO $ of $1+i\R$ obtained by applying the map $z\to -z$ to $\mcO$ and translating by $1$. By uniqueness of solutions of \eqref{28X17.6c} we have
$$
 \phi_{[\gamma _{\lambda}]}=\phi_{[\hat \gamma _{\lambda}]}
 \,,
$$
as both solutions have the same boundary data.
Equivalently, for all $\lambda\in[0,1]$ and $s\in \R$, the solution $\phi(\tau,x)$ satisfies
$$
 \phi\big(1-t_\lambda(s)- i \varphi_\lambda(s),x\big) =
  \phi \big(t_\lambda(s) + i \varphi_\lambda(s),x\big)
   \,.
$$
In particular
$$
 \phi(1-t - i 0,x ) = \phi(t + i 0,x)
  \ \mbox{ for} \ t\in (-\frac12-\eta, \frac 12 + \eta)
 \,,
$$
and extendability of $\phi^0$ to a function which is defined for all $t\in \R$ and periodic in $t$ follows.

We have just shown that the period of the solution is at least that of $\psi^0$. Clearly, a smaller period of the solution would imply a smaller period of the boundary data. Hence the periods in the interior and at the boundary coincide.

\subsection{On P3.}
The solutions constructed so far will be complex-valued in general (which will be fine if complex scalar fields are considered, in which case we are done). If real-valued solutions are sought, we need to
justify P3 above. For this suppose that, in addition to \eqref{5XI17.8}, the complex conjugate of $\psi$ satisfies
$$
 \overline{\psi(\overline\tau,x)} = \psi(\tau,x)
 \,.
$$
Then the functions $(t,x)\mapsto \psi(t+i0,x)$ and  $\varphi\mapsto \psi(0+i\varphi)$ are  real-valued. Further,
\begin{equation}\label{5XI17.15}
   \overline{\phi(\overline \tau,x)}
\end{equation}
is holomorphic in $\tau$, solves the same equation,
with same boundary data. By uniqueness
$$
 \overline{\phi(\overline\tau,x)} = \phi(\tau,x)
 \,,
$$
and real valuedness of $\phi^0$ readily follows.

\bigskip

In order to summarise what has been done so far, it is convenient to introduce a definition:

\begin{definition}
  \label{D13XI17.1} Let $\mcV$ be open.
A function $\psi(t+i\varphi,x)$, $x\in \mcV$ will be called \emph{admissible} if it is meromorphic in $\tau\equiv t+i\varphi \in \C$ at fixed $x$,  satisfies
\bel{6XI17.0}
  \psi(\tau ,x) = \psi(-\tau ,x) = \psi (\tau +p + i q,x)
 = \overline{\psi(\overline \tau ,x)}
  \,,
  \quad
  \forall \  p,q \in \Z
   \,, \ x \in \mcV
\,,
\ee
%
with all singularities lying on the union of vertical lines $\Re \tau  = \frac 12 + \Z$. We further assume that there exists $\eta>0$ so that $\psi$ is smooth in all variables on $\mcU\times \mcV$, where $\mcU$ contains the image of $\{(t,\varphi)\in [-1/2,1/2]\times\R\}$ by the map \eqref{11XII17.1+}.
\end{definition}

Existence of many non-trivial admissible functions is established in Appendix~\ref{A6XI17.1}.

We have proved:

\begin{theorem}
  \label{T13XI17.1}
  Let $(\Omega,g)$ be a smooth compact Riemannian manifold with smooth boundary and let $V$ be a smooth strictly positive function on $\Omega$.
  Suppose that $F$ is real analytic, real valued, and satisfies $F(0)=0$.
Given any smooth admissible function $\psi(\tau,x)$, $x\in \partial\Omega$, there exists $\epsilon_0>0$ such that for all $0<\epsilon<\epsilon_0$ the boundary-value problem
$$
 \left\{
   \begin{array}{ll}
 \Box_\glorentz \phi = F(\phi), & \hbox{$(t,x)\in \R\times \Omega$;} \\
      \phi(t,x)= \epsilon \psi(t+i 0,x), & \hbox{$(t,x)\in \R\times \partial\Omega$,}
   \end{array}
 \right.
$$
has a global, smooth, real-valued, time-periodic solution.
\end{theorem}

\section{Einstein equations with a negative cosmological constant}
  \label{s13XI2017}

In the presence of a negative cosmological constant the proof of existence of solutions of the Einstein equations, possibly with matter fields, is essentially a repetition of the proof of Theorem~\ref{T13XI17.1}, with some technicalities thrown in. We will present a formal statement and proof only in vacuum, and comment on the non-vacuum solutions in Remark~\ref{R19XI17.1} below.

It is convenient to start with some terminology and notation. We work in space-time dimension $d:=n+1$ and we normalise the cosmological constant to
\bel{10III17.21}
 \Lambda=-\frac{n(n-1)}2
  \, .
\ee

Let $M$ be a compact $n$-dimensional manifold with boundary.
Let us denote by
\begin{equation}\label{19XI17.1}
 (\rho,x^A)
\end{equation}
local coordinates near $\partial M$.
A Lorentzian metric $\glorentz$ on $\R\times M$ will be called \emph{conformally compactifiable} if the metric $\rho^2 \glorentz$ can be $C^k$-extended across $\R\times\partial M$ while  preserving signature.


Similar definitions apply to Riemannian metrics $\griem$, which in our case will be defined on $S^1\times M$, with conformal boundary at infinity $S^1 \times \partial M$. In this case the coordinate along $S^1$ will be denoted by $\varphi$, and we will use the notation
\begin{equation}\label{19XI17.1+}
 (\rho,x^a)\equiv (\rho, \varphi, x^A)
\end{equation}
for the local coordinates near the conformal boundary $S^1\times \partial M$.

We will say that a tensor field $\psi\equiv\psi_{ab}dx^a dx^b$ defined on $S^1\times \partial M$ is admissible if
the tensor components $\psi_{ab}$ are admissible in the sense of Definition~\ref{D13XI17.1}, with $z=t+i \varphi$ and $\mcV= \partial M$ there.
If needed such tensors can, and will, be extended to a neighborhood of $S^1\times \partial M$ in $S^1\times M$ by requiring $\partial_\rho \psi_{ab}=0$.

Let $\znabla$ denote the covariant derivative associated with a  two-covariant, symmetric, non-degenerate, possibly complex valued tensor field $\zgriem$. Set
\bel{10+XI17.2}
 \lambda^\mu:= \frac 1 {\sqrt{\det \griem}} \znabla_\alpha ({\sqrt{\det \griem}}  \griem^{\alpha \mu})
 \,,
\ee
where the derivative $\znabla $ is  understood as a covariant derivative operator acting on tensor densities.
Let us denote by $\Rriem^\mu{}_{\alpha\beta\gamma}$ a tensor field obtained by using the usual formula for the Riemann tensor of $\griem$, similarly for the Ricci tensor $\Rric$, we set
\bel{10+XI17.1}
 \Rric^H_{\alpha\beta}:= \Rric_{\alpha\beta}
  +
  \frac 12 \big(
 \griem _{\alpha\mu} \znabla_\beta \lambda^\mu
 +
 \griem_{\beta\mu} \znabla_\alpha \lambda^\mu
  \big)
 \,.
\ee
Then the linearisation with respect to the metric, at $\griem= \zgriem$, in dimension $d=n+1$, of the map
$$
 \zgriem \mapsto \Rric^H_{\alpha\beta} +  (d-1) \griem
$$
is the operator
\bel{10+XI17.5}
 \zP:=\frac 12(\zDelta_L+2n)
 \,,
\ee
where the \emph{Lichnerowicz Laplacian} $\zDelta_L$  acts on   symmetric
two-tensor fields $h$ as
\bel{10+XI17.4}
 \zDelta_Lh_{\alpha \beta}:=-\znabla^\gamma\znabla_\gamma h_{\alpha \beta}
  +\zRric_{\alpha \gamma }h^\gamma {_\beta}+\zRric_{\beta\gamma }h^\gamma {_\alpha }-2\zRriem_{\alpha \gamma \beta \delta }h^{\gamma \delta}
  \,.
\ee
We will say that a metric $\zgriem$ is {\it non-degenerate} if $\zDelta_L+2n$ has no
$L^2$-kernel.

Large classes of non-degenerate Einstein
metrics are described in~\cite{Lee:fredholm,ACD2,mand1,mand2,ChDelayKlingerNonDegenerate,Gursky}.

We have:

\begin{theorem}\label{T19XI17.1}
 Let $n=\dim M\ge 3$,
and
 consider a static Lorentzian real-valued Einstein metric $\zglorentz$ of the form
\beal{8+XI17.1}
 &\zglorentz  = -\zV^2 dt ^2 + \underbrace{\zg_{ij}dx^i dx^j}_{=:\zg}\, ,
 &
\\
 &
 \partial_t \zV =   \partial_t \zg=0
 \, ,
 &
\eeal{8+XI17.2}
where $\zV$ is strictly positive.
Assume that the associated Riemannian
 metric
\beal{8+XI17.3}
 &
  \zgriem  =  \zV^2 d\varphi^2 +  \zg
 &
\eea
is non-degenerate.
For every  time-periodic admissible time-periodic tensor field $[\psi_{ab}dx^adx^b]$ on $\R\times \partial M$ there exists  $\epsilon_0>0$ so that for all $0<\epsilon<\epsilon_0$   there exists a
time-periodic, with the same period,
near to $\zglorentz$, stationary Lorentzian real-valued vacuum metric
such that, in local coordinates near the
conformal boundary $\partial M$,
\bel{8+XI17.3+}
 \glorentz - \zglorentz = \rho^{-2} \big(\epsilon\psi_{ab} + O (\rho)\big) dx^a dx^b
 \,.
\ee
\end{theorem}

\medskip

\begin{Remark}
  \label{R20XI17.2}
The anti-de Sitter metric $\glorentz$ satisfies the hypotheses of Theorem~\ref{T19XI17.1}, whence existence of many periodic vacuum solutions near the anti-de Sitter metric.
\qed
\end{Remark}
%
%

\begin{Remark}
  \label{R20XI17.2+}
Uniqueness holds within the class of solutions constructed, up to multiplication of $\psi$ by a periodic conformal factor, but uniqueness in the class of time-periodic solutions is not clear.
\qed
\end{Remark}
\begin{Remark}
  \label{R20XI17.2+a}
The construction here applies to the Riemannian Einstein metrics obtained by a ``Wick rotation'' for a large subset (possibly all except the critical spherical ones, compare~\cite{ChDelayKlingerBH}) of the Birmingham family~\cite{Birmingham} of metrics. In the static or stationary case this leads to Lorentzian solutions with a smooth event horizon. On the other hand, the doubly periodic, in $t+i\varphi$, solutions constructed here are unlikely to  lead to  Lorentzian metrics with well behaved horizons.
\qed
\end{Remark}

\begin{Remark}
  \label{R19XI17.1}
As already mentioned, the argument applies to Einstein equations with a large class of matter sources, e.g.\  for Yang-Mills-Higgs-Einstein-Maxwell-Chern-Simons-dilaton-scalar field-$f(R)$ equations. The generalisation, which should be obvious to the reader, proceeds by applying the implicit function theorem to the whole system of equations and following the arguments presented in the proof below; the relevant isomorphism theorems have been established in~\cite{ChDelayKlinger,ChDelayKlingerBH}. The free asymptotic data for each of the fields have been described in~\cite{ChDelayKlinger} in the stationary case. The periodic solutions are obtained by replacing the time-independent free data there by sufficiently small, time-periodic, admissible in the sense of Definition~\ref{D13XI17.1}, holomorphic fields.
Taking non-trivial periodic data for one (or all) of the  matter fields together with
time-independent conformally flat asymptotic data for the metric leads to non-trivial time-periodic solutions with a standard cylindrical boundary at infinity.
\qed
\end{Remark}

\begin{Remark}
  \label{R19XI17.5}
The solutions, whether vacuum or with matter fields, will have the usual Fefferman-Graham-type expansions at the conformal boundary. The analysis in Section~7 of~\cite{ChDelayKlinger} of finiteness of total energy of solutions, or that of their total mass in the case of a conformally flat conformal boundary, applies verbatim to the time-periodic solutions constructed here.
\qed
\end{Remark}

\begin{Remark}
  \label{R25XI17.1}
The period of the solutions can be chosen arbitrarily, in the proof we chose it equal to one for definiteness. The proof guarantees that, both here and in Theorem~\ref{T13XI17.1}, the period of the solution will be one when the  boundary data are one-periodic, but does not guarantee a shorter period of the solution if the Lorentzian boundary data happen to have a shorter one.
\qed
\end{Remark}

\noindent{\sc Proof:}
The result is obtained through a mixture of the methods of Section~\ref{s28X17.1} and of the constructions in~\cite{GL,Lee:fredholm}.

The calculations underlying the back-and-forth transitions from the hyperbolic equations satisfied by a real-valued real-analytic Lorentzian metric to an elliptic equation satisfied by a complex-valued symmetric tensor field  can be found in Appendix~\ref{A7XI17.1}.

Let $\tau=t+i\varphi$ be a complex variable as in Section~\ref{s28X17.1}. Let us denote by $x$ the local coordinates on $M$, and note that the Riemannian metric $\zgriem$, defined on $S^1\times M$, associated with $\zglorentz$, depends only upon $x$: $\mcL_{\partial_\varphi} \zgriem = 0$. Let us continue to denote by $\psi$ any extension of the boundary-data tensor $\psi$ from $S^1\times \partial M$ to the interior of $S^1\times M$. We wish to construct a symmetric complex valued tensor field  $\griem$, holomorphic in $\tau $, such that
\bel{8+XI17.3+a}
 \griem (t+i\varphi, x) - \zgriem (x)  = \rho^{-2} \big(\epsilon\psi_{ab}(t+i\varphi, x) + o (\rho^2)\big) dx^a dx^b
 \,.
\ee

It is routine to check that Theorem~A of \cite{Lee:fredholm} applies to boundary conformal classes with sufficiently small imaginary part, leading to non-degenerate symmetric two-covariant tensors with Riemannian real part and small imaginary part on $S^1\times M$ which solve the Einstein equation with negative cosmological constant $\Lambda$. (A more detailed treatment in a similar context can be found in~\cite{ChDelayKlingerBH}.) It then follows from this theorem that for any $0<t_0 < 1/2$  there exists $\epsilon_0>0$  so that for all $0<\epsilon<\epsilon_0$  and  for any $ t \in (-t_0,t_0)$ there exists a solution of the Einstein equations with boundary data (in the sense of \eqref{8+XI17.3+a})
$$
 S^1\times \partial M \ni ( \varphi,   x^A) \mapsto  \epsilon \psi_{ab}(  t + i \varphi, x^A)
 \,.
$$
Let us denote by $\griem_{ t}$ this solution. Standard results show that the tensor fields $\griem_t$ are jointly smooth in all variables
$$
 (t,\varphi, x)\in (-t_0,t_0)\times S^1\times M
 \,.
$$

We claim that the family $  t \mapsto \griem_{  t}$ of solutions so obtained is holomorphic in $t + i  \varphi$. For this, let us denote by $P(  t )$ the operator obtained by  linearising the ``harmonically-reduced Riemannian Einstein equations'',
\bel{10III17.8a}
  \Rric^H  +n \griem  =0
  \,,
\ee
at $\griem=\griem_{ t}$. The operator $P(0)$ has no $L^2$-kernel by hypothesis, which therefore remains true by continuity for all $  t\in [-t_0,t_0]$, after reducing   $\epsilon_0$  if necessary. Differentiating \eqref{10III17.8a} with respect to $\overline{ \tau}$ one has
\bel{10III17.8b}
 0=\partial_{\overline {  \tau}} ( \Rric^H  +n \griem)
 = P(  t) \partial_{\overline {  \tau}} \griem_{  t}
  \,.
\ee
This provides an elliptic equation for $\partial_{\overline { \tau}} \griem_{ t}$, with asymptotic data $\epsilon \partial_{\overline { \tau}}  \psi\equiv 0$, which implies
\bel{10III17.8c}
 \partial_{\overline {  \tau}} \griem_{ t} \equiv 0
  \,.
\ee
Hence the tensor fields
$$
 \griem (  t+ i   \varphi,  x  ):= \griem_{  t}(  \varphi, x )
$$
are indeed holomorphic in $\hat \tau$, as claimed.

Now, since $\psi(\tau,\cdot)=\psi(-\tau,\cdot)$ by hypothesis, the fields $\griem (\tau,x)$ have the same asymptotic data as  $\griem (-\tau,x)$. By uniqueness,
\begin{equation}\label{22XI17.5}
 \griem (  \tau,x)= \griem (- \tau,x)
 \,.
\end{equation}

Next, the complex-conjugate fields $\overline {\griem (t+i\phi,x)}$ satisfy the same equations as $ \griem (t+i\phi,x)$, with boundary data
$$
 \overline {\psi  (t+i\phi,\cdot)} = \psi(t-i\varphi, \cdot)
 \,.
$$
By uniqueness of solutions,
$$
 \overline {\griem  (t+i\phi,x)} = \griem(t-i\varphi, x)
 \,.
$$
Equivalently,
$$
 \overline {\griem  (\overline \tau ,x)} = \griem(\tau, x)
 \,.
$$
This, together with \eqref{22XI17.5}, implies that the fields $\griem$ are real-valued on both the real and the imaginary axes of the $\tau$-plane.

As the next step, we wish to show that there exists an analytic extension of $\griem(\tau, x)$ from the strip $\tau\in (-t_0,t_0)\times \R$, obtained so far, to a whole neighborhood of the real axis.
For this we will use  the biholomorphic map $\chi$ of \eqref{11XII17.1+} to
 define a new holomorphic variable on the boundary $S^1\times \partial M$,
$$
 \tau = t+ i\varphi \mapsto   \hat \tau\equiv \hat t+ i \hat \varphi :=  \chi^{-1}(t+i\varphi)
 \,,
$$
together with new coordinates on $S^1\times\partial M$,
$$
(x^a) \equiv ( \varphi,  x^A) \mapsto     (\hat x^a)\equiv (  \hat \varphi,\hat x^A) := (  \hat \varphi,x^A)
 \,.
$$
We define a new boundary-data tensor
\begin{equation}\label{20XI17.3}
  \hat \psi(\hat \tau, \hat x):= \psi_{ab}\big(\chi^{-1}(\hat \tau,x)\big) \frac{\partial x^a}{\partial \hat x^c} \frac{\partial   x^b}{\partial \hat x^d} d \hat x^c d \hat x^d
\end{equation}
(compare \eqref{17XI17.1+} below).
The tensor field $\hat \psi $ is defined on $(-1/2,1/2)\times\R$, holomorphic in $\hat \tau$ there, satisfies
$$
 \hat\psi(-\hat \tau)=\hat\psi(\hat \tau)
 = \overline{\hat\psi(\overline{\hat{\tau}})}
 \,,
$$
and is periodic in $\hat \varphi$: for all $p\in \Z$:
$$
 \hat \psi \big(\hat \tau + i ( \hat \varphi+p),x\big)= \hat \psi (\hat \tau +i \hat \varphi,x)
 \,.
$$
Note that for any $\eta>0$ the sets
$$
 \mbox{$
 \chi\big((-1/2,1/2)\times\R\big)$ and  $1+\chi\big((-1/2,1/2)\times\R\big)$}
$$
overlap, forming a $\C$-neighborhood of the interval $[-1,1]$ lying on the real axis (cf.\ Figure~\ref{fig1}, where the relevant interval lies on the imaginary axis). There exists $\eta_0>0$ so that $\chi\big([-1/2,1/2]\times\R\big)$ avoids all singularities of $\psi(\tau,\cdot)$ for all $0\le\eta\le\eta_0$. This avoidance property remains true for $1+\chi\big([-1/2,1/2]\times\R\big)$, as follows from the symmetry properties of $\chi$ and $\psi$.

There exist $0<\epsilon_1\le \epsilon_0$ and $\eta_1>0$ so that for all $0<\epsilon\le \epsilon_1$, $0<\eta\le \eta_1$ and $\hat t \in [-1/2,1/2]$ we can again solve the harmonically-reduced Einstein equations to obtain a family of metrics, which  will be denoted by $\hat \griem_{\hat t} (\hat \varphi, x)$, which asymptote to
\begin{equation}\label{20XI17.5}
 \rho^{-2}(d\rho^2 + \Big(\zgriem _{ab}(x)
  + \epsilon \psi_{ab}\big(\phi^{-1}(\hat \tau,x)\big)
    \Big)\frac{\partial x^a}{\partial \hat x^c}
        \frac{\partial   x^b}{\partial \hat x^d} d \hat x^c d \hat x^d
\end{equation}
as $\rho$ tends to infinity.

One shows as before
\bel{10III17.8c+}
 \partial_{\overline {\hat \tau}} \hat \griem_{\hat t} \equiv 0
\,,
\quad
 \hat \griem (\hat \tau,x)= \hat \griem (-\hat \tau,x) = \overline{
 \hat \griem (\overline{\hat \tau},x)
}
  \,,
\ee
where of course $\hat \griem (\hat t+ i \hat \varphi,x):= \hat \griem_{\hat t}(\hat \varphi, x)$.

Since the hypotheses on the functions $\psi_{ab}$ imply reality on both the real and the imaginary axes (cf.\ Appendix~\ref{A6XI17.1}),  the asymptotic data \eqref{20XI17.5} for $\hat\griem_0(\hat \varphi,   x)$ have vanishing imaginary part. The map $\chi$, when restricted to the imaginary axis $\Re \tau=0$, provides a diffeomorphism from $\R$ to $\R$ for all small parameters $\eta$ there, in which case \eqref{20XI17.5} is a standard (real-valued) change of coordinates $\varphi\mapsto \hat \varphi$ at $\tau=0+i\varphi$ for a (usual, real-valued) Riemannian metric.

Consider the tensor field
\begin{equation}\label{22XI17.11}
  \hat \griem_{\mu\nu}(\chi(\tau),x) \frac{\partial \hat x^\mu }{\partial x^\alpha}
        \frac{\partial  \hat x^\nu}{\partial x^\beta} dx^\alpha d x^\beta
 \,.
\end{equation}
It has the same asymptotic data as $\griem(\tau,x)$ at $\tau=0+i\varphi$ by construction, therefore coincides with $\griem(\tau,x)$ there. By uniqueness of analytic extensions it coincides with $\griem (\tau,x)$ wherever both are defined. It provides the desired analytic extension of $\griem(\tau,x)$ to a neighborhood of the interval $t\in [-1/2-\eta,1/2+\eta]\subset \C$.

Periodicity in $t$ follows now as in the scalar field case.
\qed
%

\appendix

\section{Remarks on doubly periodic holomorphic functions}
 \label{A6XI17.1}

The aim of this appendix is to exhibit large classes of functions as needed in our existence results, namely: Meromorphic functions $f$, satisfying
\bel{6XI17.0+}
  \overline{f(\overline z)}= f(z) = f(-z) = f(z+p + i q)
  \,,
  \quad
   \forall \ p,q \in \Z
   \,,
\ee
%
analytic in a neighborhood of the  axes, with all singularities lying on the union of vertical lines $\Re z = \frac 12 + \Z$.

For this, for $n\in \N$, $n\ge 2$ and $a\in (0,1)$ consider the series
\begin{equation}\label{6XI17.1}
 f_{n,a}(z) = \sum_{p,q\in \Z} \frac 1 {(z- \frac 12 - i a - p - i q )^n}
 \,.
\end{equation}
These series converge and define indeed a family of meromorphic functions such that
\begin{equation}\label{6XI17.2}
 \mbox{$ f(z+p+iq)=f(z)$ for all $p,q\in \Z$,}
\ee
with singularities located on the set $\{\frac 12 + ia + p + iq\,,\ p,q\in\Z\}$.

Functions satisfying \eqref{6XI17.2} will be referred to as \emph{doubly periodic}. Any finite linear combination of the functions of the form  \eqref{6XI17.1} with $a=a_k$ for some $a_k \in \R$  will be meromorphic, doubly periodic,
with singularities located on the union of the singular sets $\{\frac 12 + ia _k+ p + iq\,,\ p,q\in\Z\}$.

Consider, next any doubly periodic meromorphic function $f$. Then
\begin{equation}\label{6XI17.3-}
  f_+(z):= \frac 12 \big(f(z) + f(-z)
   \big)
\end{equation}
is again meromorphic, doubly periodic, and satisfies
\begin{equation}\label{6XI17.3}
  f_+(z)= f_+(-z)
   \,.
\end{equation}
Functions satisfying \eqref{6XI17.3} will be referred to as \emph{even}.
Note that even analytic functions have only even powers of $z$ in their Taylor-series expansion at the origin.

Let $f$ be again a doubly periodic even meromorphic function and set
\begin{equation}\label{6XI17.4}
  f_r(z):= \frac 12 (f(z)+ \overline{f(\overline z)})
\end{equation}
Then $f $ is again doubly periodic, meromorphic, even, and for $z$ such that $\overline z=z$ this reduces to $f_r(z)= \Re f(z)$, whence $f_r$ is real on the real axis. It is also real on the imaginary axis for even $f$ since then
$$
 z = \Im z
 \ \Rightarrow
 \
 \overline z = - z
 \ \Rightarrow
 \
 f_r(z) = \frac 12 (f(z)+ \overline{f(- z)})
 = \frac 12 (f(z)+ \overline{f( z)})
  =
   \Re f(z)
  \,.
$$

We conclude that for any finite collection $(n_\ell ,a_\ell ,\alpha_\ell )$, $\ell=1,\ldots,N$, $n_\ell \ge 2$, $n_\ell \in \N$, $a_\ell \in(0,1)$, $\alpha_\ell \in \C$, the functions
\begin{equation}\label{6XI17.5}
  f =\sum_{\ell=1}^N \big((\alpha_\ell  f_{n_\ell ,a_\ell })_+\big)_r
\end{equation}
satisfy all the requirements set forth at the beginning of this section.

\section{``Complex coordinate transformations''}
 \label{A7XI17.1}

Consider a Lorentzian metric tensor
\begin{eqnarray}
  \phantom{xxxx}
   \glorentz
   &=&
    \glorentz_{\mu\nu}(t+i 0 ,x)dx^\mu dx^\nu
 \label{8XI17.1}
 \\
 \nn
   &=& \glorentz_{00}(t+i 0 ,x) dt^2
 +2 \glorentz_{0 j }(t+i0,x)dt dx^j
 +\glorentz_{k\ell}(t+i0,x) dx^k dx^\ell
  \,,
\end{eqnarray}
which is real-analytic in a variable $t$, so that we can analytically extend the metric coefficients $g(t,x)$ to functions $\glorentz_{\mu\nu}(t+i\varphi,x)$ of a complex variable $\tau=t+i\varphi$ to a neighborhood of the real axis $\Im \tau =0$.
Let $I$ be an nonempty open interval containing the origin. We wish to show that for $t\in I$ the ``substitution''
\begin{equation}\label{7XI17.1}
  (t+i0 ,dt,\partial_t)\mapsto (0+i\varphi , id\varphi, -i \partial_\varphi)
\end{equation}
in the metric tensor and its derivatives
maps the Ricci tensor of $\glorentz$ to the \emph{formal Ricci tensor}
of the (possibly complex valued) tensor field
$$
 \griem= - \glorentz_{00}(0+ i \varphi,x) d\varphi^2
 +2 i  \glorentz_{0 j }(0+i\varphi,x)d\varphi dx^j
 + \glorentz_{k\ell}(0+i \varphi,x) dx^k dx^\ell
  \,,
$$
with
$\varphi\in J \subset \R$ for some nonempty open interval $J$ containing $0$.
Here, by \emph{formal Ricci tensor of $g$} we mean the tensor calculated using the usual formulae for the components of the Ricci tensor associated with a real-valued symmetric non-degenerate tensor field $g$.

For this, given a smooth curve
$$
 \R \supset I \ni s\mapsto \gamma(s)=t(s)+i\varphi(s)\in \C
  \,,
$$
we consider the complex valued-tensor field
\begin{eqnarray}\label{17XI17.1}
  \griemg (s,x^i)
   & := &
      \glorentz_{00}({\gamma(s)},x)\big(\frac{d{\gamma(s)}}{ds}\big)^2ds^2
\\
 \nonumber
 &&
 +2  \glorentz_{0 j }({\gamma(s)},x) \frac{d{\gamma(s)}} {ds} dx^j ds
 +\glorentz_{k\ell}( {\gamma(s)} ,x) dx^k dx^\ell
  \,.
\end{eqnarray}
Setting  $(y^\mu)\equiv (y^0,y^i):=(s,x^i)$, \eqref{17XI17.1} can be rewritten as
\begin{equation}\label{17XI17.1+}
  \griemg_{\alpha\beta}=   \glorentz_{\mu\nu}
   \frac{\partial x^\mu}{\partial y^\alpha}
   \frac{\partial x^\nu}{\partial y^\beta}
  \,,
\end{equation}
which takes the same form as the transformation law of a tensor field under a change of coordinates. Defining ${\griemg}^{\alpha\beta}$ to be the matrix inverse to ${\griemg}_{\alpha\beta}$, we define the Christoffel symbols $ \RGamma^\alpha{}_{\beta \gamma}$ of ${\griemg}$ using the usual formula:
\begin{equation}\label{17XI17.2}
  \RGamma^\alpha{}_{\beta \nu} = \frac 12 {\griemg}^{\alpha\mu}
   \big(
    \partial_{y^\beta} {\griemg}_{\mu \nu}
   +
    \partial_ {y^\nu} {\griemg}_{\beta\mu}
   -
    \partial_ {y^\mu}{\griemg}_{\beta \nu}\big)
    \,.
\end{equation}
In view of our hypothesis, that the metric functions are holomorphic in $\tau$, one has
(compare \eqref{5XI17.11})
$$
 \partial_s \glorentz_{\alpha \beta}(\gamma(s),x) =  {\partial_s \gamma} \partial_\tau \glorentz_{\alpha \beta}(\gamma(s),x)
 \,.
$$
This implies that the  $\RGamma^\alpha{}_{\beta \gamma}$'s can be obtained from the Christoffel symbols $\mathbf{\Gamma}^\alpha{}_{\beta\gamma}$ of $\glorentz$ by the usual transformation formula
\bel{Contrla}
  \RGamma^\mu {}_{\nu \rho }=
    \mathbf{\Gamma}^\lambda {}_{\kappa \sigma } \frac{\partial
    y^\mu }{\partial{x^\lambda }}\frac{\partial
    x^\kappa}{\partial{y^\nu }}\frac{\partial x^\sigma }{\partial{y^\rho }}
 +
  \frac{\partial y^\mu }{\partial{x^\lambda }}
\frac{\partial^2 x^\lambda }{\partial{y^\rho }\partial{y^\nu }}
 \,.
 \ee
A standard calculation implies now that the field
\bel{15II.11}
   \Rriem^\delta{}_{\gamma \alpha \beta }   = \partial_{y^\alpha } \RGamma^\delta _{\beta \gamma }-  \partial_{y^\beta}  \RGamma^\delta{}_{\alpha \gamma }
+  \RGamma_{\alpha \sigma}^\delta { \RGamma^\sigma_{\beta \gamma } } -   \RGamma_{\beta \sigma}^\delta  \RGamma^\sigma_{\alpha \gamma }
 \,.
\ee
which we will call the \emph{Riemann tensor} of $\griem$, can be obtained from the Riemann tensor $\mathbf{R}^{\delta }{}_{\gamma  \alpha  \beta  }$ of $\glorentz$ using the  transformation law of tensors:
\bel{15II.11+}
   \Rriem^\delta{}_{\gamma \alpha \beta }   =  \mathbf{R}^{\delta'}{}_{\gamma' \alpha' \beta' }
    \frac{ \partial y^{\delta}} {\partial x^{\delta'}}
    \frac{ \partial x^{\gamma'}} {\partial y^\gamma }
    \frac{ \partial x^{\alpha'}} {\partial y^\alpha }
    \frac{ \partial x^{\beta'} }{\partial y^ \beta}
 \,.
\ee
Similarly for the contraction $\Rriem^\delta{}_{\gamma \delta \beta } $, which we called the Ricci tensor.
Our claim follows by taking $\gamma(s)= i s$.

\bigskip

\noindent{\sc Acknowledgements} Useful discussions with Peter Aichelburg, Erwann Delay, Robin Graham, Paul Klinger and Maciej Maliborski are acknowledged. The research of PTC was supported in
part by the Austrian Science Fund (FWF), Project  P29517-N27, and by the Polish National Center of Science (NCN) under grant 2016/21/B/ST1/00940. The author is grateful to the Fields Institute for hospitality and support during an essential part of work on this paper.

\bibliographystyle{amsplain}

\bibliography{../references/newbiblio,%
../references/reffile,%
../references/bibl,%
../references/prop,%
../references/hip_bib,%
../references/newbib,%
../references/PDE,%
../references/netbiblio,%
Periodic-minimal,references}
\end{document}